\documentclass[prl,twocolumn,aps,letterpaper, preprintnumbers,superscriptaddress]{revtex4}
\usepackage{hyperref}
\usepackage{verbatim}
\usepackage{amsmath}
\usepackage{latexsym}
\usepackage{revsymb}
\usepackage{yfonts}
\usepackage{ifthen}
\usepackage{tcolorbox}
\usepackage{natbib}
\usepackage{amsfonts}
\usepackage{amsmath}
\usepackage{amssymb}
\usepackage{amsthm}
\usepackage{graphicx}
\usepackage{bm}
\usepackage{bbm}
\usepackage{epsfig,color,amssymb}
\usepackage{subfigure}
\usepackage{amsfonts}
\usepackage{amscd}
\usepackage{amsmath}
\usepackage{multirow}
\usepackage{chemarrow}
\usepackage{dcolumn}
\usepackage{bm}
\usepackage{graphicx}
\usepackage{enumerate}
\usepackage{epsfig}
\usepackage{subfigure}
\usepackage{xcolor}
\usepackage{multirow}
\usepackage{ulem}
\usepackage{braket}
\usepackage{comment}
\usepackage{enumitem}
\usepackage{amsthm}

\renewcommand{\emph}[1]{\textit{#1}}

\begin{document}

\title{Nondiffracting Supertoroidal Pulses: Optical ``K\'arm\'an Vortex Streets''}


\author{Yijie Shen}\email{y.shen@soton.ac.uk}
\author{Nikitas Papasimakis}
\affiliation{Optoelectronics Research Centre \& Centre for Photonic Metamaterials, University of Southampton, Southampton SO17 1BJ, United Kingdom}
\author{Nikolay I. Zheludev}
\affiliation{Optoelectronics Research Centre \& Centre for Photonic Metamaterials, University of Southampton, Southampton SO17 1BJ, United Kingdom}
\affiliation{Centre for Disruptive Photonic Technologies, School of Physical and Mathematical Sciences and The Photonics Institute, Nanyang Technological University, Singapore 637378, Singapore}
\date{\today}


\begin{abstract}
Recently introduced supertoroidal light pulses [Nat. Commun. 15, 5891 (2021)] are a family of space-time nonseparable freespace electromagnetic excitations with unique topological properties including skyrmionic field configurations, fractal-like patterns of singularities, and extended areas of energy backflow. Here we report nondiffracting supertoroidal pulses (ND-STPs), propagation-robust skyrmionic and vortex-ring field that endure the singular configurations over arbitrary propagation distances. Intriguingly, the field structure in of ND-STPs has a strong similarity with a von K\'arm\'an vortex street, a pattern of swirling vortices observed in fluid and gas dynamics that is responsible for the ``singing'' of suspended telephone lines in wind. We argue that ND-STPs are of interest as directed energy channels for telecom applications. 
\end{abstract}

\maketitle

The topological properties of spatial light fields have been a subject of fascination and intense research interest over the last half century~\cite{nye1974dislocations,berry2000making,dennis2009singular} with implications for light-matter interactions~\cite{bao2021light,ozawa2019topological,malinauskas2016ultrafast}, nonlinear physics~\cite{luu2015extreme,krogen2017generation,shen2017gain}, spin-orbit coupling~\cite{rego2019generation,dorney2019controlling,shen2019optical}, microscopy and imaging~\cite{davis2020ultrafast,pu2021unlabeled,pu2020label}, metrology~\cite{yuan2019detecting}, and information transfer~\cite{xie2018ultra,qiao2020multi,wan2021divergence}. Light pulses can also be simultaneously structured in space and time domains~\cite{chong2020generation,bliokh2021spatiotemporal,kondakci2017diffraction,yessenov2019weaving,zdagkas2021observation}. Among them, the toroidal pulses are of particular interest as they can engage toroidal excitations in matter~\cite{papasimakis2016electromagnetic,raybould2017exciting,zdagkas2021observation} and exhibit space-time nonseparability and isodiffraction~\cite{zdagkas2020space,shen2021measures,shen2022nonseparable}. 
As a recent generalization of toroidal pulses, the supertoroidal pulses (STPs) also exhibit striking topology including fractal-like singularities, vortex rings, energy backflows, and optical skyrmionic patterns~\cite{zdagkas2019singularities,shen2021supertoroidal}.

In this Letter, we show that the transverse divergence of STPs can be tuned to archive nondiffracting supertoroidal pulses (ND-STPs). Intriguingly, the structure of an ND-STP resembles that of K\'arm\'an vortex street (KVS), staggered vortex arrays observed in fluid and gas dynamics~\cite{1954aerodynamics}, also previously observed in CW structured light fields~\cite{scheuer1999optical,molina2002observation,molina2002singular,shen2018vortex}.

\begin{figure}[t!]
	\centering
	\includegraphics[width=\linewidth]{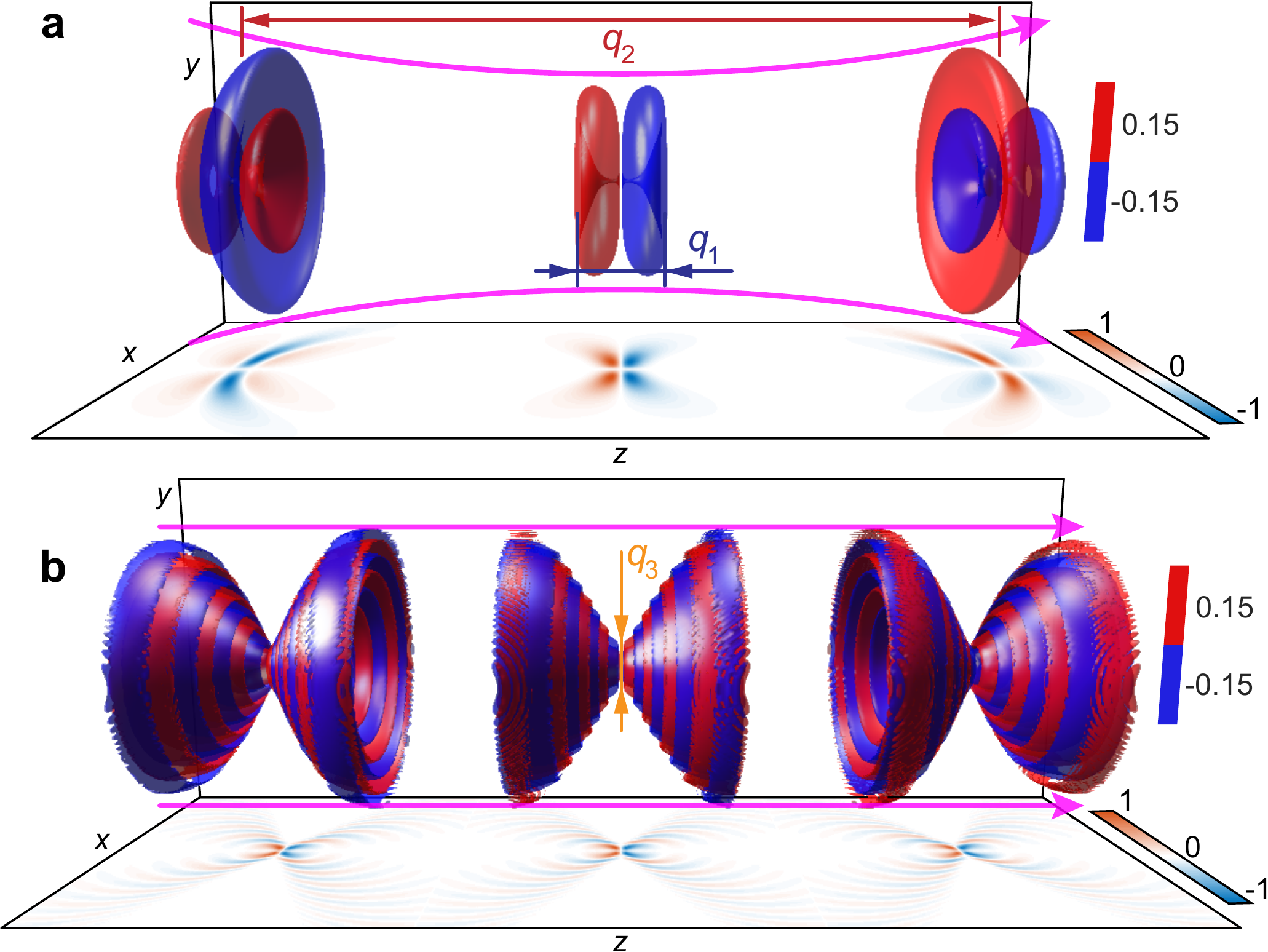}
	\caption{An elementary toroidal pulse~\cite{hellwarth1996focused} (\textbf{a}) and a ND-STP (\textbf{b}): the isosurfaces with $\pm0.15$ amplitude level and the amplitude distributions on $x$-$z$ plane at various times $t=0,\pm q_2/(2c)$. Blue, red, and yellow arrows mark optical cycle ($q_1$), Rayleigh range ($q_2$), and transverse divergence ($q_3$), respectively. Parameters in simulation: $q_2/q_1=40$ for the toroidal pulse; $q_2/q_1=100$ and $q_3/q_1=1$ for the ND-STP. Purple arrows mark the propagation profile. See Supplementary Video~1 for the dynamic evolutions.} 
	\label{f1}
\end{figure}

\begin{figure*}[t!]
	\centering
	\includegraphics[width=\linewidth]{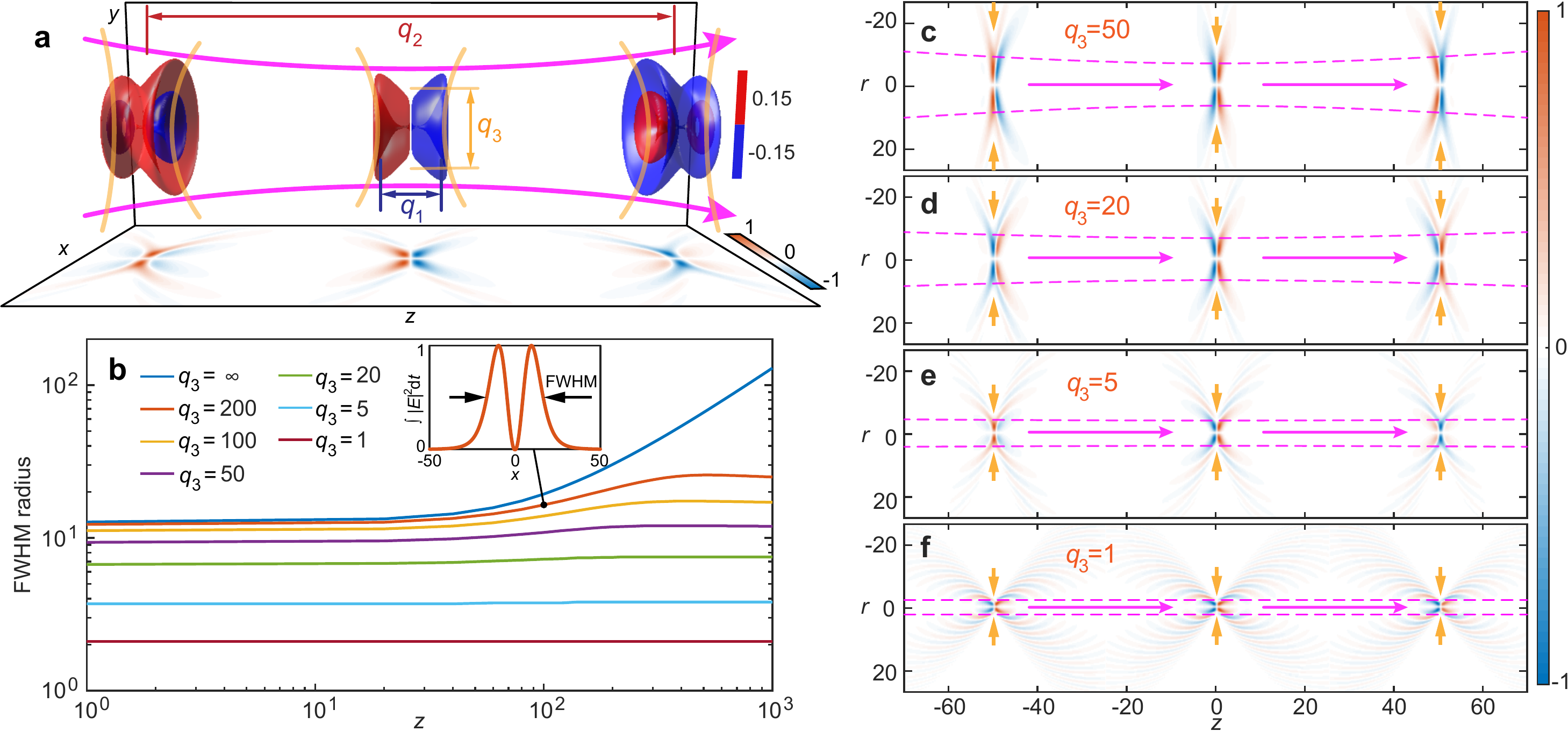}
	\caption{(\textbf{a}) A characteristic example of a transversely divergent pulse with $q_3/q_1=20$, which is represented by electric field amplitude isosurfaces at levels $\pm 0.15$ of the maximum field amplitude in three different moments in time {$t=-q_2/(2c),0,q_2/(2c)$}. The presented case lies between the fundamental toroidal pulse [$q_3\to\infty$; see Fig.~Fig.~\ref{f1}(\textbf{a})] and a ND-STP [$q_3=q_1$; see Fig.~\ref{f1}(\textbf{b})]. The transition from the former to the latter can be seen in Supplementary Video~1. (\textbf{b}) The evolution of pulse width in the transverse upon propagation for different $q_3$ values. The inset shows a 1D cross-section of the pulse intensity in the transverse plane, where the pulse width is quantified by the full width at half maximum (FWHM). \textbf{c-f}, The spatiotemporal evolution of transversely divergent pulses with various $q_3$ values of 100, 20, 5, and 1, respectively. In each panel, the purple dashed lines mark the FWHM of the corresponding pulses. Unit of length: $q_1$.} 
	\label{f2}
\end{figure*}

STPs and ND-STPs originate from the ``electromagnetic directed-energy pulse trains'' (EDEPT) introduced by Ziolkowski~\cite{ziolkowski1989localized}. Following Ziolkowski's theory~\cite{ziolkowski1989localized}, the solutions are derived through a scalar generating function $f(\mathbf{r},t)$ fulfilling the scalar wave equation, $( \nabla^2-\frac{1}{{{c}^{2}}}\frac{{{\partial }^{2}}}{\partial {{t}^{2}}})f\left( \mathbf{r},t \right)=0$, where $\mathbf{r}=(r,\theta,z)$ represents spatial cylindrical coordinate, $t$ is time, $c=1/\sqrt{\varepsilon\mu}$ is the speed of light, and the $\varepsilon$ and $\mu$ are the permittivity and permeability of medium. Here $f(\mathbf{r},t)=f_0{e}^{-s/{q_3}}(q_1+i\tau)^{-1}(s+q_2)^{-\alpha}$, where $f_0$ is a normalizing constant, $s=r^2/(q_1+i\tau)-i\sigma$, $\tau=z-ct$, $\sigma=z+ct$, $q_1,q_2,q_3$ are real positive parameters with units of length, and the real dimensionless parameter $\alpha$ must
satisfy $\alpha\ge1$, to ensure finite energy of the pulse~\cite{ziolkowski1989localized,hellwarth1996focused}. In order to retrieve the electromagnetic fields, we construct the Hertz potentials $\mathbf{\Pi}(\mathbf{r},t)=\mathbf{\hat{n}}f(\mathbf{r},t)$, where $\mathbf{\hat{n}}$ can be arbitrary vector operator, and then the electromagnetic fields of transverse electric (TE) mode can be obtained  by~\cite{ziolkowski1989localized,hellwarth1996focused}: $\mathbf{E}(\mathbf{r},t)=-{{\mu }_{0}}\frac{\partial }{\partial t}\bm{\nabla} \times \mathbf{\Pi }$, $\mathbf{H}(\mathbf{r},t)=\bm{\nabla} \times \left(\bm{\nabla}\times \mathbf{\Pi } \right)$. If the Hertz potential is set as a transverse vortex vector field, i.e. $\mathbf{\hat{n}}=\bm{\nabla} \times\mathbf{\hat{z}}$, the resulting electromagnetic field is an azimuthally polarized toroidal pulse~\cite{hellwarth1996focused}, as shown in Fig.~\ref{f1}(\textbf{a}). The transverse magnetic (TM) mode can be obtained by exchanging electric and magnetic fields of the TE mode. If the Hertz potential is defined as $\mathbf{\hat{n}}=\mathbf{\hat{x}}$, the result is a linearly polarized pancake-like pulse (polarized along $y$-axis)~\cite{feng1999spatiotemporal}.
In prior works, the conditions $q_1\ll q_2$, $\alpha=1$, and $q_3\to\infty$ were assumed to generate various focused structured pulses~\cite{hellwarth1996focused,feng1999spatiotemporal,feng1998gouy,lekner2004localized,lekner2004helical}, where $q_1$ and $q_2$ determin the wavelength and the longitudinal divergence or Rayleigh length, $z_0=q_2/2$ of the pulses, respectively, as marked in Fig.~\ref{f1}(\textbf{a}). The case of $\alpha\geq 1$ and $q_3\to\infty$ was recently studied leading to the introduction of supertoroidal pulses~\cite{shen2021supertoroidal}. In this work, we explore STPs with finite $q_3$ (see Supplementary Material for detailed derivation). A characteristic example of such a pulse with $q_3=q_1$ and $\alpha=1$ is plotted in Fig.~\ref{f1}(\textbf{b}). Here, $q_3$ defines the degree of transverse divergence: with decreasing value of $q_3$ the pulse envelope is gradually squeezed into a dumbbell-like shape and eventually becomes nondiffracting for $q_3=q_1$ and $\alpha=1$, fulfilling the definition of nondiffraction $|E(r,\theta,z,t)|=|E(r,\theta,z+\Delta z,t+\Delta z/v)|$~\cite{hernandez2013non}, where $\Delta z$ is a given propagation distance and $v$ is group velocity of the wave here $v=c$, see Supplementary for the proof. 

\begin{figure*}[t!]
	\centering
	\includegraphics[width=\linewidth]{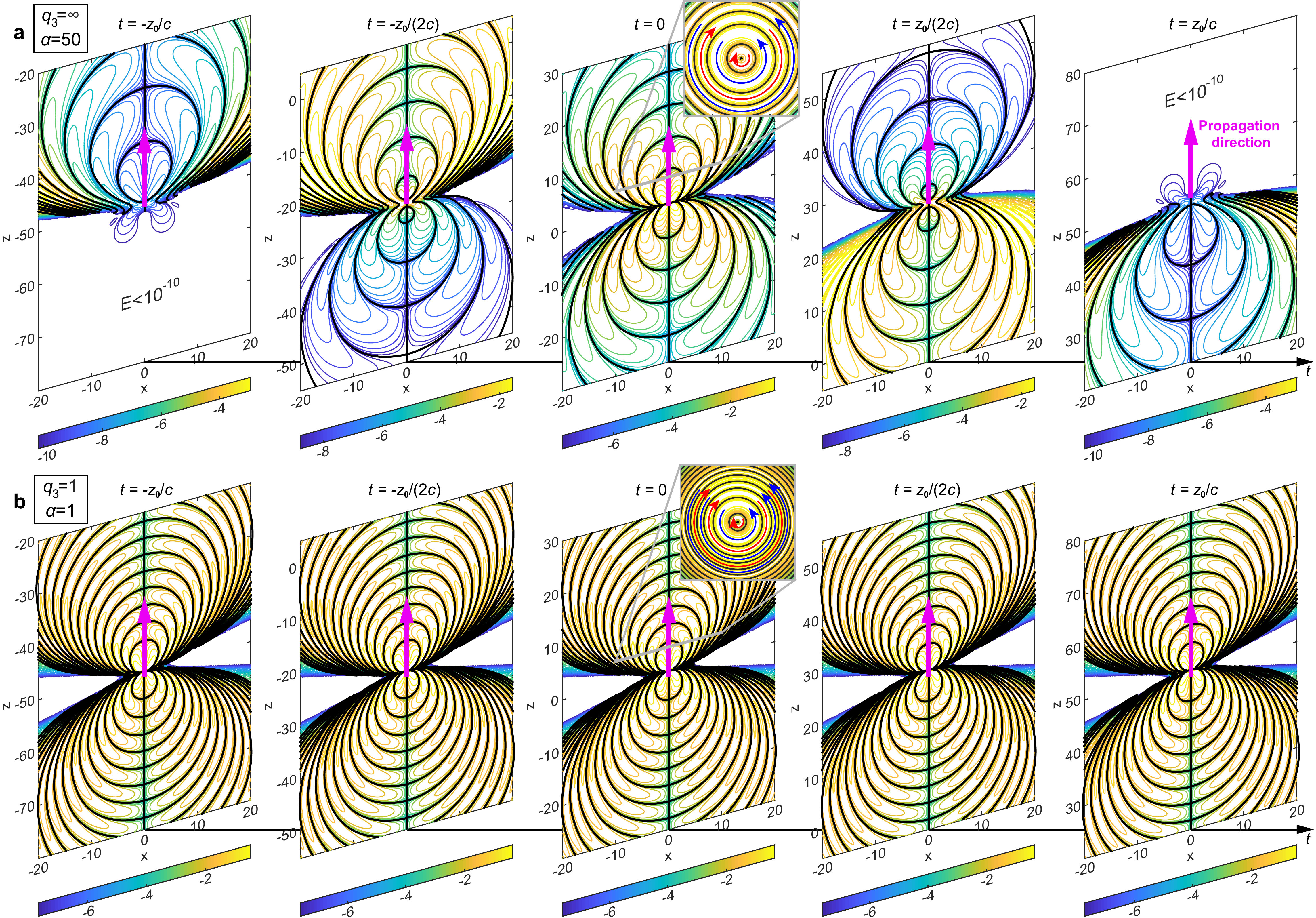}
	\caption{The singular structures of electric field evolving with propagation versus the time for a focused supertoroidal pulse (\textbf{a}) and a ND-STP (\textbf{b}): each panel presents an isoline plot of the logarithm of the electric field of the pulse in the $x$-$z$ plane at a given time. The singularities are highlighted by bold black lines. Unit of length: $q_1$. See Supplementary Video~2 for the dynamic evolutions of figures (\textbf{a}) and (\textbf{b}), and other general cases with different parameters, respectively.}
	\label{f3}
\end{figure*}

The evolution of the pulse from diffracting to nondiffracting is illustrated in Fig.~\ref{f2}. An intermediate case of a weakly-diffracting supertoroidal pulse in terms of $q_1$, $q_2$, and $q_3$, is presented Fig.~\ref{f2}(\textbf{a}). To illustrate the dependence of the (non)diffracting nature of the pulse on $q_3$, we calculate the $z$-dependent full width at half maximum (FWHM) radius of the transverse intensity pattern of the STPs with $q_2=100q_1$ and $q_3$ varying from infinity to $q_1$, for propagation distances from focus to $z=10^3q_1$, see Fig.~\ref{f2}(\textbf{b}). In the extreme case of $q_3\to\infty$ (fundamental toroidal light pulse), the FWHM of the pulse follows a hyperbolic trajectory similar to a conventional focused Gaussian beam \cite{hellwarth1996focused}. With  decreasing $q_3$, the divergence becomes weaker and the pulse approaches a nondiffracting state for $q_3\le5q_1$. Figures~\ref{f2}(\textbf{c})-\ref{f2}(\textbf{f}) show the spatiotemporal evolution upon propagation for toroidal pulses with different $q_3$ values. 
Here, decreasing $q_3$, results in a faster spatiotemporal evolution of the cycle structure of the pulse (see Supplementary Video~1), due to its increasingly complex shape. When the $q_3$ value is decreased further to $q_1$, then the pulse becomes X-shaped. 
Note that $q_3$ can be an arbitrarily small positive real number and can take values smaller than $q_1$, in which case the pulse remains nondiffracting and becomes further squeezed. Figure~\ref{f1}(\textbf{b}) plots the special case of $q_3=q_1$, where the X-shape reveals the conical structure in space, accommodating multiple cycles along the conical surface. Upon propagation, the cycle structure keeps evolving on the conical surface akin to a breather (see Supplementary Video~1). 
Note that similar X-type nondiffracting pulses have been considered previously both theoretically and experimentally and are typically termed Bessel-X pulses~\cite{saari1997evidence,besieris20052+,bowlan2009measuring}. However, previous works focused on scalar long-pulses within the slowly-varying amplitude envelope approximation. Here, our ND-STPs are few-cycle, space-time nonseparable with nontrivial electromagnetic toroidal topology lacking in the Bessel-X pulses. A typical example involves the intriguing skyrmions and KVS structures is discussed in this work. 

The emergence of ND-STPs allows us to explore intriguing topological optical effects. In our recent work, we showed that supertoroidal pulses exhibit a complex topology (controlled by the parameter $\alpha$), including self-similar fractal-like patterns, matryoshka-like singularity arrays, skyrmions, and areas of energy backflow~\cite{shen2021supertoroidal}. Here, we show that such topological structures are also present in nondiffracting supertoroidal pulses. However, whereas in the former case, the self-similar topological structures were only instantaneously observed at focus, in the ND-STP, they can be observed over arbitrarily long distances [see example in Fig.~\ref{f3}(\textbf{b})]. For instance, we have shown in our previous work that STPs exhibit a complex topology that evolves rapidly upon pulse propagation with elements of self-similarity consisting of concentric matryoshka-like spherical shells over which the electric field vanishes [see Fig.~\ref{f3}(\textbf{a})]. Similar propagation-robust fractal-like topological structures also exist in the ND-STP [see Fig.~\ref{f3}(\textbf{b})] and persist upon pulse propagation (see also Supplementary Video~2).

\begin{figure}
	\centering
	\includegraphics[width=\linewidth]{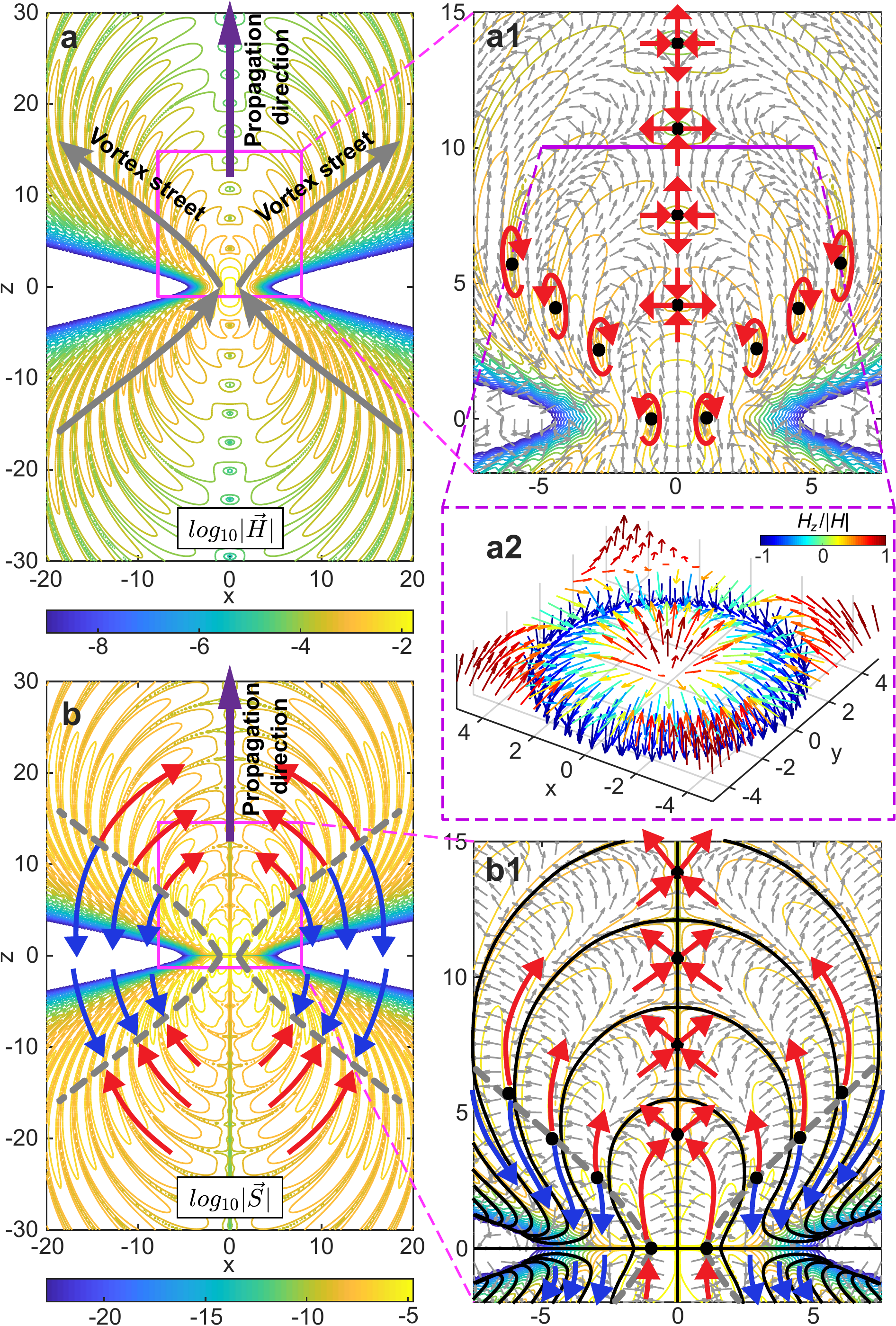}
	\caption{Magnetic field (\textbf{a}) and Poynting vector (\textbf{b}) distributions of a ND-STP with $q_2/q_1=100$ and $q_3/q_1=1$ at $t=0$. The contours show the logarithm of the modulus of the corresponding vectors. (\textbf{a1}) A zoom-in of the magnetic field with the arrow plot showing the vector distribution, the black dots referring to the singularities, and surrounding red arrows marking the type of the vector singularities (vortex and saddle) serving as guide to the eye. (\textbf{a2}) Arrow plot showing a skyrmionic structure of magnetic field in the transverse ($x$-$y$) plane at $z=10$ (indicated by the purple solid line in (\textbf{a1})). (\textbf{b1}) A zoom-in of the Poynting vector field, where the arrow plot corresponds to the vector distribution, solid black lines and dots mark the zeros of the Poynting vector, red and blue arrows indicate areas with forward and backward energy flow, respectively. Unit of length: $q_1$.}
	\label{f4}
\end{figure}

The complex topology of the ND-STPs manifests also in the magnetic field distribution. Figure~\ref{f4}(\textbf{a}) shows the magnetic field distribution of a ND-STP with $q_2/q_1=100$ and $q_3/q_1=1$ at $t=0$. The magnetic field includes both radial and longitudinal components resulting in vector singularities of vortex and saddle types. The saddle-type singularities are distributed along the propagation axis, while the vortices trace an off-axis trajectory, see Fig.~\ref{f4}(\textbf{a1}). Such a singularity distribution induces multiple electromagnetic skyrmions at transverse planes of the pulse, see Fig.~\ref{f4}(\textbf{a2}). In contrast to the propagating electromagnetic skyrmions in prior works that only exist around 
beam focus and collapse rapidly upon propagation~\cite{shen2021supertoroidal,shen2021topological,shen2022generation}, here in the ND-STP, the skyrmions persist upon propagation with their topological texture exhibiting a periodic behaviour alternating between four different skyrmion types (combination of opposite two polarities and two helical angles), see Supplementary Video~3.

\begin{figure}[t!]
	\centering
	\includegraphics[width=\linewidth]{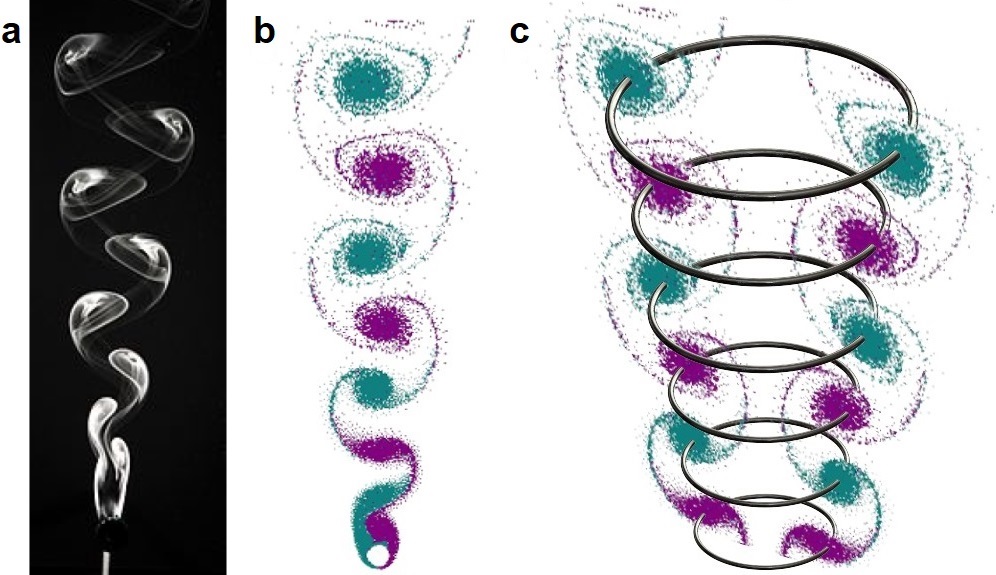}
	\caption{(a) Visualization of a KSV behind a cylinder in air, where the flow is visualized by releasing glycerol vapour in the air near the cylinder. (b) Simulation of a 2D KSV, where the right- and left-handed vortices were highlighted by purple and cyan, respectively. (c) Artistic impressions of a vortex-ring street analogue to the structure of magnetic field of a ND-STP.} 
	\label{f5}
\end{figure}

Figure~\ref{f4}(\textbf{b}) shows the Poynting vector field distribution of the ND-STP, which possesses layered energy forward-flow and backflow structures, see Fig.~\ref{f4}(\textbf{b1}). Interestingly, energy transport is mediated by the vortex arrays of the magnetic field. Vortices on the front half of the pulse act as energy sources, whereas the vortices in the rear half behave as sinks. In between sources and sinks, we observe areas of extended backflow. In areas of high intensity, energy flows primarily forward, ensuring the propagation of the pulse.

Moreover, we observe that the vortex arrays exhibited by the ND-STPs form a striking trail of two-vortex clusters (with opposite circulations, namely a vortex dipole) propagating in a periodic staggered manner, evocative of the KVS structure [Fig.~\ref{f5}(\textbf{a})]. In fluid dynamics, a KVS is such a classic pattern of swirling vortices caused by a nonlinear process of vortex shedding, which refers to the unsteady separation of flow of a fluid around blunt bodies. Vortex-street-like optical fields were reported previously in stationary optical fields exhibiting phase vortex patterns~\cite{scheuer1999optical,molina2002observation,molina2002singular,shen2018vortex}. In contrast, here we observe KVS-like structures in propagating electromagnetic spatiotemporal pulses in linear regime. In fluid dynamics, a KVS is a pattern of repeating swirling vortices constructed in the flow velocity field, whilst in our optical analogy, the KVS pattern is constructed by the electric or magnetic field of ND-STPs. Moreover, whereas KVS in fluid dynamics is a nonlinear phenomenon, here we show that similar effects can be observed in a linear system. The analogy between KVS in fluid flows and ND-STPs can be drawn further by considering for instance the motion of electrons along the vortex streets of a TM ND-STP or the propagation of supertoroidal pulses in nonlinear media. Finally, note that while the KVS is usually demonstrated in 2D [Fig.~\ref{f5}(\textbf{b})], here the KVS in ND-STP corresponds to a 3D set of vortex rings, due to the cylindrical symmetry [Fig.~\ref{f5}(\textbf{c})].

ND-STPs are vector polarized, space-time nonseparable pulses and as such they can be generated by spatially gradient metasurfaces, similarly to the toroidal pulses discovered by Helwarth and Nouchi \cite{hellwarth1996focused, zdagkas2021observation}. Of particular importance here is the relation between the size of the generating metasurface (aperture) and the distance over which the generated pulses propagate without diffraction. We have shown that in contrast to the well known Bessel beams, the field of the ND-STPs is exponentially confined (see Supplementary). We argue that the latter is less sensitive to the metasurface size and thus practical implementations of the latter remain closer to their ideal form.

In conclusion, we demonstrate nondiffracting toroidal pulses. Importantly, the sophisticated vector field configuration of ND-STPs induce robust topological structures including fractal-like singularities, skyrmions, vortex rings, and energy backflows, all of which can stably propagate at an infinitely long distance. Due to its propagation-robust topology, the ND-STP acts as a spectacular display of striking trail of staggered electromagnetic vortex dipole arrays with stable propagation, evocative of the classic KVS. The ND-STPs and optical KVSs unveil intriguing analogies between fluid transport and flow of energy in structured light. In particular, the robust topological structure of ND-STPS that remains invariant upon propagation could be used for long-distance information transfer encoded in the topological features of the pulses. Finally, we anticipate that ND-STPs will inspire potential applications such as light-matter interactions, superresolution microscopy, telecommunications, remote sensing and lidar.

\bibliographystyle{naturemag}

\textbf{Acknowledgments.} The authors acknowledge the supports of the MOE Singapore (MOE2016-T3-1-006), the UKs Engineering and Physical Sciences Research Council (grant EP/M009122/1, Funder Id: http://dx.doi.org/10.13039/501100000266), the European Research Council (Advanced grant FLEET-786851, Funder Id: http://dx.doi.org/10.13039/501100000781), and the Defense Advanced Research Projects Agency (DARPA) under the Nascent Light Matter Interactions program.

\bibliography{sample}

\end{document}